\documentclass{aa}

\usepackage{txfonts}
\usepackage{graphicx}	
\usepackage{amsmath}	
\usepackage{amssymb}	
\usepackage{newtxtext}
\usepackage[varvw]{newtxmath} 
\usepackage{array}
\usepackage{float}
\usepackage{gensymb}
\usepackage[colorlinks=true,linkcolor=blue,citecolor=blue, urlcolor=blue]{hyperref}
\bibpunct{(}{)}{;}{a}{}{,}
\usepackage{multirow} 
\usepackage{fontawesome}  

\begin{document}


\title{Accelerating galaxy dynamical modeling using a neural network for joint lensing and kinematics analyses}

\titlerunning{Accelerating dynamical modeling}

\author{
         Matthew R. Gomer\inst{1}\fnmsep     \thanks{\email{mgomer@uliege.be}}
         \and
         Sebastian ~Ertl\inst{2,3}
         \thanks{\email{ertlseb@mpa-garching.mpg.de}}
         \and
         Luca~Biggio\inst{4} 
         \and
         Han ~Wang\inst{2,3}
         \and
         Aymeric Galan\inst{3,2,5}
         \and
         Lyne ~Van de Vyvere\inst{1}
         \and
         Dominique ~Sluse\inst{1}
         \and
         Georgios ~Vernardos\inst{5,6,7}
         \and
         Sherry H.~Suyu\inst{3,2,8}
}
%
\institute{
    STAR Institute, Quartier Agora - All\'ee du six Ao\^ut, 19c B-4000 Li\`ege, Belgium 
    \and 
    Max-Planck-Institut f\"ur Astrophysik, Karl-Schwarzschild Str. 1, 85748 Garching, Germany
    \and
    Technical University of Munich, TUM School of Natural Sciences, Department of Physics, James-Franck-Stra\ss{}e~1, 85748 Garching, Germany
    \and
    Eidgen\"ossische Technische Hochschule Z\"urich - 
    CH-8092 Z\"urich, Switzerland
    \and
    Institute of Physics, Laboratory of Astrophysics -
    Ecole Polytechnique Fédérale de Lausanne (EPFL),
    1290 Versoix, Switzerland
    \and
    Department of Astrophysics, American Museum of Natural History, Central Park West and 79th Street, NY 10024, USA
    \and
    Department of Physics and Astronomy, Lehman College of the City Unuversity of New York, Bronx, NY 10468, USA
    \and
   Academia Sinica Institute of Astronomy and Astrophysics (ASIAA), 11F of ASMAB, No.~1, Section 4, Roosevelt Road, Taipei 10617, Taiwan
    }

\abstract{Strong gravitational lensing is a powerful tool to provide constraints on galaxy mass distributions and cosmological parameters, such as the Hubble constant, $H_0$. Nevertheless, inference of such parameters from images of lensing systems is not trivial as parameter degeneracies 
can limit the precision in the measured lens mass and cosmological results. External information on the mass of the lens, in the form of kinematic measurements, is needed to ensure a precise and unbiased inference. Traditionally, such kinematic information has been included in the inference after the image modeling, using spherical Jeans approximations to match the measured velocity dispersion integrated within an aperture. However, as spatially resolved kinematic measurements become available via IFU data, more sophisticated dynamical
modeling is necessary. Such kinematic modeling is expensive, and constitutes a computational bottleneck which we aim to overcome with our Stellar Kinematics Neural Network (SKiNN). SKiNN emulates axisymmetric modeling using a neural network, quickly synthesizing from a given mass model a kinematic map which can be compared to the observations to evaluate a likelihood. With a joint lensing plus kinematic framework, this likelihood constrains the mass model at the same time as the imaging data. We show that SKiNN's emulation of a kinematic map is accurate to considerably better precision than can be measured (better than $1\%$ in almost all cases). Using SKiNN speeds up the likelihood evaluation by a factor of $\sim 200$. This speedup makes dynamical modeling economical, and enables lens modelers to make effective use of modern data quality in the JWST era.}


\maketitle

\section{Introduction}
Gravitational lensing is a powerful tool that can measure the mass distributions of galaxies, and even offers a method to measure the Hubble parameter, $H_0$, which is independent of the distance ladder \citep{Refsdal64}. In this context, the lens is typically an early-type galaxy (ETG) with a time-variable source, from which time delays between the multiple images can be used alongside a model of the lensing potential to provide a measure of distance. However, lensing degeneracies can introduce systematic uncertainties in the lens mass distribution
which must be accounted for to recover an accurate $H_0$. The most critical of such degeneracies is the Mass Sheet Degeneracy \citep[MSD,][]{Falco85}, which expresses a specific transformation of the mass distribution that leaves all imaging observables invariant, but affects the time delays. This leads to a measure of $H_0$ which is dependent on the choice of model. One must turn to external information to break the degeneracy and decide which model to keep. One such form of external information is the stellar kinematics of the lens galaxy, which must be measured and modeled in conjunction with the lensing model if one wishes to derive an accurate value of $H_0$ \citep{Treu02a,Treu02b,Koopmans03}.

The historically established method to use kinematics to break the MSD has consisted of combining lens models with a single aperture measurement of the galaxy velocity dispersion
\citep[e.g.,][]{Suyu10a,Sonnenfeld12,Wong17,Birrer19,Rusu20}. First, a lens model is performed, which provides a mass model and light model of the lens galaxy. From this model, one can use spherical Jeans  approximations \citep{Binney87} to estimate a predicted aperture velocity dispersion. Joining these observations together is performed by comparing the predicted velocity dispersion value to the observed value, and combining this kinematic $\chi^2$ together with the lens model likelihood to determine the parameters which match both observations with the maximum total likelihood. Typically this kinematic constraint is included in post-processing, with only a single aperture constraint to help decide between the already converged lensing-inferred results near the maximum of the imaging likelihood. The main limitation of this approach is that the velocity dispersion is not considered jointly with the lens model to help guide the sampling of the likelihood. Typically, a single aperture measurement only weakly favors a given lens model over another, with relatively little constraining power and relatively low contribution to the total likelihood. 

As the quality of data improves, the constraining power of kinematic information becomes more valuable. Telescopes are increasingly capable of measuring spatially-resolved velocity dispersions in galaxies. Through integral field unit (IFU) spectrographs
such as the Multi Unit Spectroscopic Explorer \citep[MUSE;][]{Bacon10}, the Keck Cosmic Web Imager \citep[KCWI;][]{Morrissey12}, and the James Webb Space Telecope (JWST) NIRSpec IFU \citep{Yildirim20}, pixelated maps of velocity measurements
for lens systems are becoming available, providing constraints on the mass over a range of radii. If combined with lens models, this information can break the MSD and improve the precision of cosmological measurements \citep{Birrer21a,Yildirim23,Shajib23}.

The challenge is that this data requires more sophisticated 
dynamical
models than the spherical Jeans models historically used for this task. Beyond spherical Jeans, the next level of generalization is to allow the model to be axisymmetric. Implementation of this model is possible using the Jeans Anisotropic Multiple Gaussian Expansion \citep[JAM;][]{Cappellari08} method, which decomposes a mass profile using Multiple Gaussian Expansion (MGE), deprojects the Gaussian components given an inclination, and calculates the $v_{\rm rms}=\sqrt{v_{\rm rot}^2+\sigma_{\rm v}^2}$ in the sky plane, where $v_{\rm rot}$ is the rotational velocity and $\sigma_{\rm v}$ is the velocity dispersion. 

JAM has been used to study the structure of nearby ETGs \citep{Cappallari11}. Spatially resolved kinematics data measured by the SAURON survey has shown that ETGs come in two distinct kinematic classes: fast rotators and slow rotators \citep{Cappellari07, Emsellem07}. Fast rotators are well-matched by an oblate axisymmetric model, and as such are well-modeled by JAM \citep{Cappellari16}. Slow rotators, meanwhile, typically have position angles of their light distributions which are misaligned with the kinematic axis, which implies they tend to have prolate or triaxial shapes \citep{Weijmans14,Loubser22}. As such, the use of JAM carries the implicit assumption that these axes are aligned for a given lens galaxy, which \citet{Krajnovic11} found to be true for approximately $90\%$ of nearby ETGs. Using JAM would provide a natural way to self-consistently model lens systems, except that the calculation is much more expensive than spherical Jeans. Combining more computationally expensive dynamical
modeling with the already-expensive lens modeling in a joint inference framework is at present prohibitive without significant computational resources. 



Current methods exist which are capable of combining spatially resolved kinematics and lensing information to varying degrees. \citet{Barnabe07} first created a joint kinematics+lensing modeling code with a goal of studying galaxy structure \citep[see also][]{Barnabe09, Barnabe12}, and as such it has not been used for $H_0$ inference. \citet{vandeVen10} self-consistently compared a lens model of imaging data with an axisymmetric kinematics model of resolved kinematic data, but the models were fit separately. 
More recently \citet{Yildirim20,Yildirim23} implemented a joint lensing and dynamics framework using JAM which is capable of time-delay cosmography, but is computationally very demanding to fit simultaneously the many lensing and kinematic measurements.

One source of inspiration for this work is that the booming field of machine learning (ML) has helped solve similar problems in related fields \citep[e.g., see review by][]{Huertas-Company23}. Neural networks (NNs) have been used to replace expensive solver operations in cosmological applications \citep{Albers19,Bonici22}, to replace stellar population synthesis in spectral modeling \citep{Alsing20}, to extract physical properties from velocity measurements of galaxies \citep{Dawson21}, and even to speed up gravitational lens modeling itself \citep{Hezaveh+17, Perreault+17, Pearson+19, Park+21, Schuldt+21, Schuldt+22, Biggio22}. This work applies a similar strategy by using a NN to emulate the expensive kinematics computation step for use within a joint lensing+dynamics
modeling framework. The NN emulates only the kinematics so as to ensure both pieces (the lens model and the emulated kinematic model) retain physical meaning independently. This strategy also allows the NN to be implementable in a modular fashion with existing lens modeling frameworks.

\begin{figure*}
    \centering
    \includegraphics[width=0.95\linewidth]{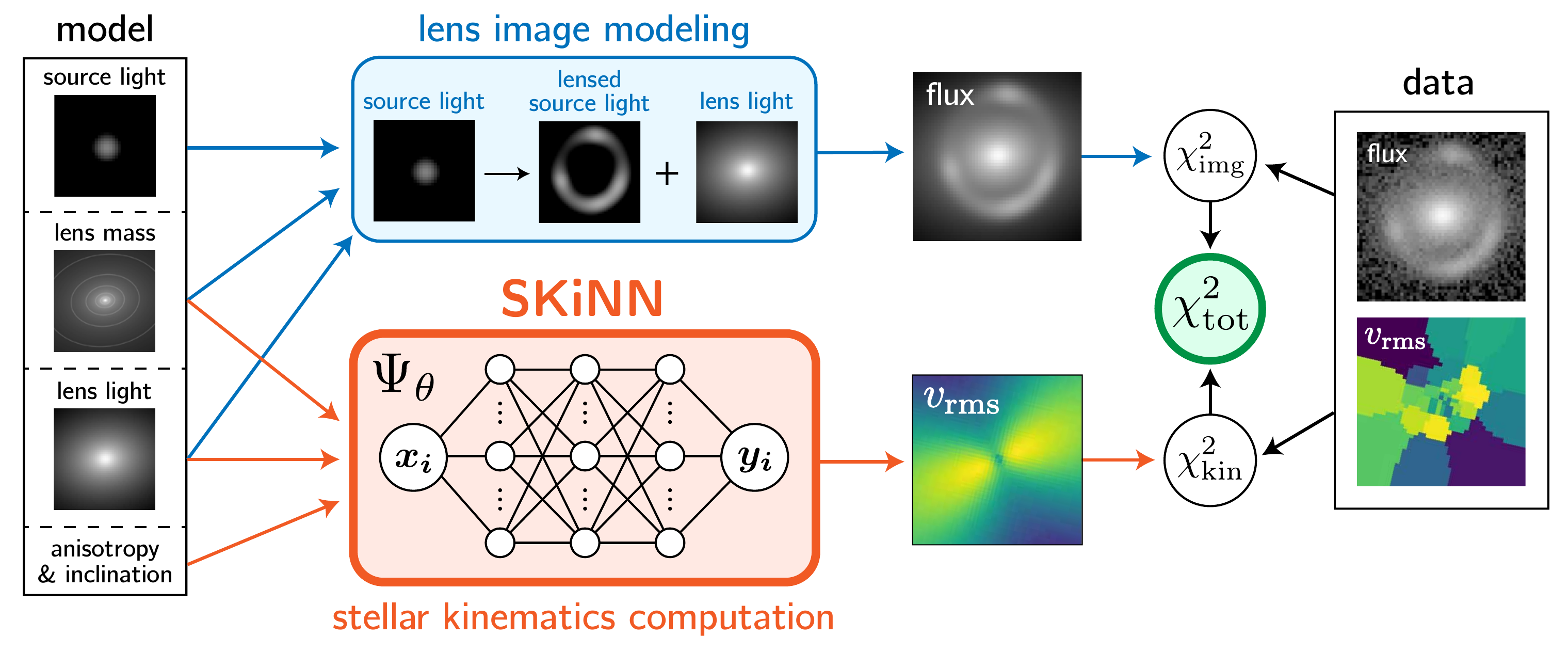}
    \caption{SKiNN's role in a joint modeling framework. SKiNN takes as an input a parametric description of the lens mass profile, lens light profile, anisotropy of the stellar orbits in the lens galaxy, and lens inclination and outputs a $v_{\rm rms}$ map used to evaluate a kinematic likelihood ($\chi^2_{\rm kin}$). This likelihood is then combined with the lensing likelihood ($\chi^2_{\rm img}$) to give a total likelihood ($\chi^2_{\rm tot}$), which is then optimized or sampled. In case of a variable lensed source with measured time delays, a time delay likelihood ($\chi^2_{\rm TD}$, not shown in the figure) can be added to the inference (see Sect.~\ref{sec:joint_modeling}).}
    \label{fig:SKiNN_scheme}
\end{figure*}

Having originally unveiled the prototype at NeurIPS 2022 \citep{NeurIPS_version}, in this work we present the Stellar Kinematics Neural Network (SKiNN), which replaces the kinematics computation for a joint lens+kinematics modeling framework. SKiNN is built to emulate JAM, producing a high-resolution velocity map of a galaxy, given a parametric description of the mass and light of the lensing galaxy as an input. Figure \ref{fig:SKiNN_scheme} shows a schematic of such a joint framework, wherein mass and light profiles are modeled through lens modeling and kinematics modeling to evaluate a joint likelihood. We highlight SKiNN's role in orange, which calculates the velocity map associated with a particular model. SKiNN is open source and available for use as a python package at \url{https://github.com/mattgomer/SKiNN}.

In this work we show that SKiNN is capable of emulating JAM to high accuracy, and does so at greatly increased speed. Although SKiNN can be incorporated into virtually any existing lens modeling code, in this work we demonstrate its usage in \texttt{lenstronomy}\footnote{
    \url{https://github.com/lenstronomy/lenstronomy}}
\citep{Birrer18, Birrer21b}. However, SKiNN is fully differentiable by construction and thus its usage is optimal within fully differentiable lens modeling codes \citep{Gu22, Galan22, Biggio22}. This work represents a proof of concept, and as such we restrict ourselves to a training set constructed via relatively simple mass and light models (see Sec. \ref{sec:SKiNN_details}), with the mindset that the SKiNN method could be expanded to more general training sets via transfer learning. The positive results obtained with SKiNN in this context suggest that generalizing its approach could pave the way for proper utilization of high-quality data expected from the next generation of telescopes.

This paper is organized with Sec. \ref{sec:JAM} reviewing the details of JAM, Sec. \ref{sec:SKiNN_details} describing creation of the training set and SKiNN architecture, Sec. \ref{sec:performance} quantifying the performance of SKiNN, and Sec. \ref{sec:joint_modeling} describing the implementation into a joint framework, Sec. \ref{sec:discussion} discussing these results, and finally concluding in Sec. \ref{sec:conclusion}.

\section{Jeans Anisotropic Multiple Gaussian Expansion (JAM)} \label{sec:JAM}

To model the kinematics of galaxies, one must implement some simplifying assumptions, chief among them being the assumption that the galaxy is in a steady state. Under this assumption, the dynamics are completely described by the gravitational potential and the 6-dimensional distribution function (DF) $f(\vec{x},\vec{v})$, describing the positions and velocities of the tracer population \citep{Cappellari16}. The problem of reconstructing the DF and the mass distribution from line-of-sight data is inherently underconstrained, worsened by the non-uniqueness of deprojection \citep{Gerhard96}. To reduce the dimensionality of the problem, a triaxial shape, axisymmetry, or spherical symmetry is often imposed.

Noting these difficulties, there are three main methods to model dynamics: Schwarzschild modeling, in which a large collection of orbits are computed to represent the whole galaxy \citep{Schwarzschild79}; N-body simulations, in which particles are simulated to match the density and other observables \citep{Syer96}; and using the stellar hydrodynamics equations developed by \citet{Jeans1922} to describe the DF within a gravitational potential.For this work, the third method is ideal to predict second-order velocity moments from a mass distribution without the expense of large simulations.

The Jeans equations are derived by multiplying the velocity moments by the collisionless Boltzmann equation, which assumes a steady-state system operating solely under the force of gravity \citep{Binney87}. Within a cylindrical coordinate system under axisymmetry, the two equations are expressed as:
\begin{equation}\label{eq:Jeans1}
    \frac{\nu \overline{v^2_R}-\nu \overline{v^2_\phi}}{R} + \frac{\partial (\nu \overline{v^2_R})}{\partial R} + \frac{\partial (\nu \overline{v_R v_z})}{\partial z} = -\nu \frac{\partial \Phi}{\partial R}
\end{equation}
\begin{equation}\label{eq:Jeans2}
    \frac{\nu \overline{v_R v_z}}{R} + \frac{\partial (\nu \overline{v^2_z})}{\partial z} + \frac{\partial (\nu \overline{v_R v_z})}{\partial R} = -\nu \frac{\partial \Phi}{\partial z},
\end{equation}where $v_R$, $v_z$, and $v_\phi$ describe the velocity components in cylindrical coordinates, $\nu$ is the tracer density (zeroth velocity moment), and $\Phi$ is the gravitational potential, using the notation that $\nu \overline{v_i v_j}=\int v_i v_j f(\vec{x},\vec{v}) {\rm d}^3 \vec{v}$.

\citet{Cappellari08} introduced JAM as a way to efficiently solve these equations. Two more assumptions are implemented: 
(1) the velocity ellipsoid is aligned with the coordinate system, and (2) 
the anisotropy $\beta_z=1-\overline{v^2_z}/\overline{v^2_R}$ is a constant spatially.
This parameter represents the degree to which stellar orbits are radially aligned, axially aligned, or isotropic. These assumptions reduce equations \ref{eq:Jeans1} and \ref{eq:Jeans2} to:
\begin{equation}
    \frac{\beta_z \nu \overline{v^2_z}-\nu \overline{v^2_\phi}}{R} + \frac{\partial (\beta_z \nu \overline{v^2_z})}{\partial R}  = -\nu \frac{\partial \Phi}{\partial R}
\end{equation}
\begin{equation}
    \frac{\partial (\nu \overline{v_z^2})}{\partial z} = -\nu \frac{\partial \Phi}{\partial z}
\end{equation}
These equations allow the tracer density and gravitational potential to yield the second-order velocity moments. From here, they must be projected given an inclination angle $i$(where $i=90^{\circ}$ is edge-on)
to give the observed velocity $v^2_{\rm rms}=v_{\rm rot}^2+\sigma_{\rm v}^2$ \citep[for projection integrals, see][]{Cappellari08}, where 
$v_{\rm rot}$
represents the observed mean LOS velocity often associated with rotation and $\sigma_{\rm v}$ represents the velocity dispersion. This inclination angle $i$ is the second parameter introduced (along with $\beta_z$) to describe an axisymmetric velocity model.

For a general mass profile, solving these equations to return a $v_{\rm rms}$ can require computationally expensive numerical integrals. However, for a two-dimensional Gaussian profile, the integral can be performed analytically. Leveraging this, JAM uses the Multiple Gaussian Expansion technique \citep[MGE,][]{Emsellem94,Cappellari02} to describe a particular profile as a sum of many elliptical Gaussian profiles, then efficiently calculates the $v_{\rm rms}$ of the whole profile by summing the contributions of the Gaussian components. Requiring the two parameters $\beta_z$ and $i$, the final result is a $v_{\rm rms}$ map which can be compared with observations.

The MGE method offers a physically realistic deprojection interpretation, producing oblate axisymmetric 3D densities. The deprojection is still not unique \citep{Rybicki87}, although the non-uniqueness has been found to have only a marginal effect on the dynamics of realistic elliptical galaxies \citep[so-called konus densities, see][]{vandenBosch97}. The effect worsens for lower inclinations, and the axisymmetric oblate interpretation is only defined if $\cos^2i<q^2$, as beyond this point the deprojected 3D axis ratio becomes imaginary. This leads to a flattest possible $q_{\rm min}$ for a given Gaussian component below which deprojection is not physical.

In summary, JAM solves the axisymmetric Jeans equations and projects the velocity distribution functions along the line of site using MGE. The end result is a map of $v_{\rm rms}$ which we will use to build the training set for SKiNN.

\section{Stellar Kinematics Neural Network (SKiNN)}\label{sec:SKiNN_details}
The goal of SKiNN is to mimic JAM and thus construct a high-resolution map of $v_{\rm rms}$ in the plane of the sky. The input for SKiNN is a list of specific values for the 8 parameters in Table \ref{table:parambounds} which describe the mass and light distributions of the lensing galaxy, as well as its inclination and anisotropy.

As a proof of concept, the present version of SKiNN is restricted to a single class of mass profiles and light profiles, although in principle the method can be expanded in the future to more general lens models (discussed in Sec. \ref{sec:limitations}). At present, SKiNN is compatible with a Power-law Elliptical Mass Distribution \citep[PEMD;][]{Barkana98} and elliptical S\'ersic light profiles \citep{Sersic63}. These profiles are widely used models in lens modeling. The PEMD mass distribution is expressed in terms of convergence (dimensionless surface density) as
\begin{equation}\label{eq:pemd}
        \kappa_{\rm PEMD}(x,y) = \theta_{\rm E}^{\gamma-1} \left(\frac{3-\gamma}{1+q_{\rm M}}\right) \left(x^2+\frac{y^2}{q_{\rm M}^2}\right)^{\frac{1-\gamma}{2}},
\end{equation}
with three parameters: $\theta_{\rm E}$ sets the mass normalization which in the circular case corresponds to the Einstein radius where $\int_0^{2 \pi}\int_0^{\theta_{\rm E}}\kappa(x,y)r dr d\phi = \pi \theta_{\rm E}^2$; $\gamma$ represents the slope of the profile with $\gamma=2$ corresponding to isothermal; and $q_{\rm M}$ represents the axis ratio of the mass distribution. The S\'ersic profile has a 2D light distribution expressed as
\begin{equation}\label{eq:sersic}
    I(x,y)=A \,{\rm exp}\left[ -k\left\{ \left( \frac{\sqrt{x^2+\left(\frac{y}{q_{L}^2}\right)^2}}{R_{\rm Sersic}} \right)^{1/n_{\rm Sersic}}-1 \right\} \right].
\end{equation}
Three parameters are relevant for the kinematics calculation: $R_{\rm Sersic}$ sets the effective radius within which half of the light is contained; $n_{\rm Sersic}$ sets the shape of the profile; and $q_{\rm L}$ represents the axis ratio of the light. The normalization $A$ does not matter for the kinematics calculation, and $k$ is a constant set to ensure the half-light property of $R_{\rm Sersic}$ \citep{Sersic63}.

\subsection{Training set construction}

Care is necessary in constructing the training set for the NN. To simply allow all parameters to take any value would waste time training over nonphysical solutions, and could worsen the accuracy over physical solutions. On the other hand, the applicability of the end product is limited to the range of the training set, so one must be sure to allow relevant parameters to vary over the whole range of interest to a modeler. We detail here our rationale for how our training set is created. 

The training set consists of a paired set of labels and images. The input label $\mathbf{x}$ is a list of 8 parameters used to describe a lens system listed in Table \ref{table:parambounds}. The image $\mathbf{y}$ is a corresponding $v_{\rm rms}$ map of the lens created using JAM as detailed further in this section.\footnote{
    JAM allows one to assign a different $q$ and different $\beta_z$ for each component of the MGE, creating the capacity to allow these quantities to change with radius. For this work, we set the values of $\beta_z$ to be constant for each training lens. While we do not directly restrict $q$ to be constant for each MGE component, the profiles we use to build the training set use a constant $q$, and as such the axis ratio for a given profile ends up not changing significantly with radius.}
Our training set consists of 4000 randomly generated pairs, our validation set uses 1000  additional random pairs, and our test set uses another 4000  additional random pairs.

We construct the training set by drawing these parameters uniformly from the ranges indicated in Table \ref{table:parambounds}. The ranges from which these parameters are drawn were chosen based on the priors for the Time Delay Lens Modeling Challenge \citep[TDLMC;][]{Ding21}, as they reflect typical properties of observed lensed quasars. We allow for a varying slope through the parameter $\gamma$ and for the light distributions to have different ellipticities than the mass through the parameters  $q_{\rm M}$ and $q_{\rm L}$. Meanwhile, we assume that the centroid positions and position angles of the mass and light align. These assumptions are based on the observed lens population, for which the mass and light models generally have mostly aligned position angles and centroid positions, but may have different axis ratios \citep{Shajib19a}.
Anisotropy is motivated by the prior used by \citet{Yildirim20}, and additionally by JAM models of SAURON observations of early-type galaxies \citep{Cappellari07}. Minimum inclination is set by the flattest Gaussian of the MGE decomposition.

\begin{table}
\centering

    \begin{tabular}{c | c |c }
        
         Parameter & Description & Training set bounds \\
         \hline
         $\theta_{\rm E}$ & Einstein radius & [0.5, 2\arcsec] \\ 
         $\gamma$ & 2D PL slope (mass) &[1.5, 2.5] \\
         $q_{\rm M}$ & Axis ratio (mass) & [0.6, 1.0] \\
         $q_{\rm L}$ & Axis ratio (light) & [0.6, 1.0] \\
         $R_{\rm Sersic}$ & S\'ersic radius (light) & [0.5$\theta_{\rm E}$, $\theta_{\rm E}$] \\
         $n_{\rm Sersic}$ & S\'ersic index (light) & [2, 4] \\
         $\beta_z$ & Anisotropy &[$-$0.4, 0.4] \\
         $i$ & Inclination & [$\arccos(0.6), 90\degree$] \\

         \hline
         Map resolution & 0.02\arcsec \\ 
         Map size & 11\arcsec \\ 

    \end{tabular}
     \caption{Settings for the creation of training sets for the NN. Parameters are sampled uniformly within the range indicated by square brackets.}
     \label{table:parambounds}
\end{table}

While the ranges of parameters are described above, numerical limitations lead us to choose a few more specifications for our training set. A singular PEMD mass distribution (i.e., with a vanishing core radius) introduces numerical effects at the lens center. The PEMD mass profile diverges in the center which, in turn, will lead to unphysical $v_{\rm rms}$ maps with very high $v_{\rm rms}$ values in the center. To avoid this, we set a core radius of $r_{\rm c}=0.08\arcsec$ which is the minimum value for which the test maps produce central $v_{\rm rms}$ values of $<500$km s$^{-1}$. We consider this a reasonable threshold for numerical effects seeing as ETGs at intermediate redshift rarely have $v_{\rm rms}$ above this value \citep[two cases in a sample of 90 galaxies;][]{Derkenne21}. In addition, we set a lower bound on the axis ratio $q_{\rm MGE}$ of the Gaussians that are determined in the MGE routine, because single Gaussian components can have a very low $q_{\rm MGE}$ although the sum of all Gaussians follows well the input light and mass profiles. Since the minimum inclination angle is set by the flattest Gaussian of the MGE decomposition, those low $q_{\rm MGE}$ Gaussians would artificially skew the distribution of inclination angles towards $90\degree$. We use $q_{\rm MGE}\in[0.6,1]$, which is the same range from which we draw the axis ratio of the mass and light for the training set.

With the parameters described above, the training set was constructed as follows. From a given input label, a PEMD mass distribution and S\'ersic light profile is defined. We then use the GLEE lens modeling software \citep[Gravitational Lens Efficient Explorer, ][]{SuyuHalkola10,Suyu12} to create the light and mass profiles which JAM requires as inputs. These profiles are then fed into JAM, which outputs the $v_{\rm rms}$ image, which we set to have a $551\times551$ pixel size at 0.02\arcsec resolution. The creation of each $v_{\rm rms}$ map takes about 21 seconds on a AMD Ryzen Threadripper 3970X CPU. Approximately 90\% of this time comes from the MGE decomposition, for which a faster method is possible in the case of constant ellipticity \citep{Shajib19b}, but JAM does not use this method in order to allow for a changing axis ratio with each MGE component.

We create $v_{\rm rms}$ maps with much higher resolution and larger field of view than typical data quality for this construction, with the intention that the map will eventually be rotated and interpolated down to data resolution. Additionally, for a general lens, the values of each pixel must be renormalized by the angular diameter distance to the lens. These steps are not necessary to evaluate the loss function of the NN, and so we will postpone their discussion until Section \ref{sec:joint_modeling}.

\subsection{Architecture}
SKiNN can be seen as a function $\Psi_{\theta}:\mathbb{R}^8\to\mathbb{R}^{d\times{d}}$, mapping an 8-dimensional vector of galaxy mass and light parameters into a $d\times{d}$ map of $v_{\rm rms}$ in the plane of the sky. Here, $\theta$ represents the set of all trainable parameters of the network. Given a training dataset $\mathcal{D}=\{\mathbf{x}_i,\mathbf{y}_i\}_{i=1}^N$ where $\mathbf{x}\in\mathbb{R}^8$, $\mathbf{y}\in\mathbb{R}^{d\times d}$ and $N$ is the size of the dataset, the training process consists in finding an optimal set of parameters $\theta^*$, such that a loss function $\mathcal{L}$, measuring the performance of $\Psi$ on $\mathcal{D}$, is minimized. In this work, the standard mean-squared-error loss is chosen:
\begin{equation}\label{mse}
    \mathcal{L} = \frac{1}{N d^2}\sum_{i=1}^N \sum^{d^2}(\Psi_{\theta}(\mathbf{x}_i)-\mathbf{y}_i)^2,
\end{equation}
which we optimize using the Adam optimizer routine \citep{Kingma14}.

While the original prototypes for SKiNN used convolutional architectures \citep{NeurIPS_version}, the current version of SKiNN is based on the Conditionally Independent Pixel Synthesis architecture \citep[CIPS,][]{cips}. Rather than using convolutions, this architecture uses the coordinates of each pixel as well as the parameter vector. 
The architecture comprises two main components: a mapping network $\mathcal{M}$ and a generator $\mathcal{G}$. The mapping network takes as input the parameter vector $\mathbf{x}$ and outputs a vector $\mathbf{w}$, called the style vector. The pixel coordinates are processed by a positional encoding $e$, which ultimately results in a significant improvement in the output image quality. The generator takes these encodings as inputs and generates the values corresponding to each pixel, where the style vector $\mathbf{w}$ is used to condition the generator by modulating its weights  as indicated in Eq. 2 in \cite{cips}. Overall, the final image is obtained by passing all the pixel coordinates as well as the parameter vector to the model, i.e. $\Psi_{\theta}(\mathbf{x}_i) = \mathcal{G}(e(\operatorname{mgrid}(d, d))| \mathcal{M}(\mathbf{x}))$, where  $\operatorname{mgrid}(d, d)$ is the meshed grid of pixel coordinates spanning from 0 to $d$ for each coordinate.
For more information about the architectural details of the model, we refer the interested reader to \citet{cips}. The CIPS architecture has been shown to produce photorealistic color images with more realistic power spectra than competing image generators such as StyleGANv2. While our output has only one $v_{\rm rms}$ channel, rather than three color channels, we find that the CIPS architecture results in improved accuracy over our earlier convolutional models. We note that the CIPS architecture requires a GPU, and as such a GPU is a hardware requirement for SKiNN.

To facilitate faster training, we exploit the symmetry of the model. Specifically, the NN is trained on only one fourth of the image ($d=226$), and when generating a full map, the output is mirrored fourfold, creating the $551 \times 551$ image. This increase in speed allows us to use the more updated CIPS architecture without a significant loss in training time. Our last step in the output is to set any negative pixels to zero. 
These negative pixels only happen on rare occasions for individual pixels in the NN output, but would be nonphysical to interpret as $v_{\rm rms}$.

\section{Performance}\label{sec:performance}
\subsection{Accuracy}
\begin{figure}
 \centering 
    \includegraphics[width=0.95\linewidth]{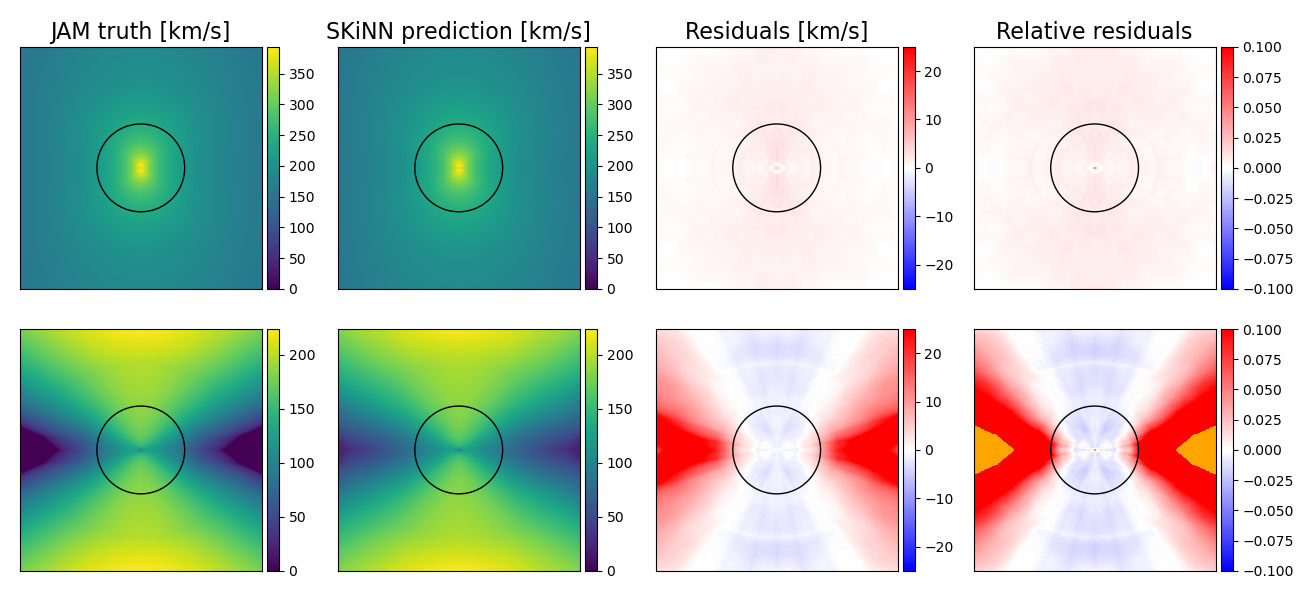}
    \caption{Accuracy of SKiNN emulation of $v_{\rm rms}$ maps in a typical case (top) and in a particularly poor case (bottom). The first two columns give the truth and predicted $v_{\rm rms}$ maps. Residual differences (third column, km s$^{-1}$) and relative residuals, evaluated as (prediction$-$truth)/truth (fourth column) are also plotted. Where pixels have zero values, relative residuals are shown in orange. The black circle corresponds to a 2\arcsec\ radius, inside which real data is most constraining.}
    \label{fig:goodbadresid}
\end{figure}

\begin{figure}
 \centering 
    \includegraphics[width=\linewidth]{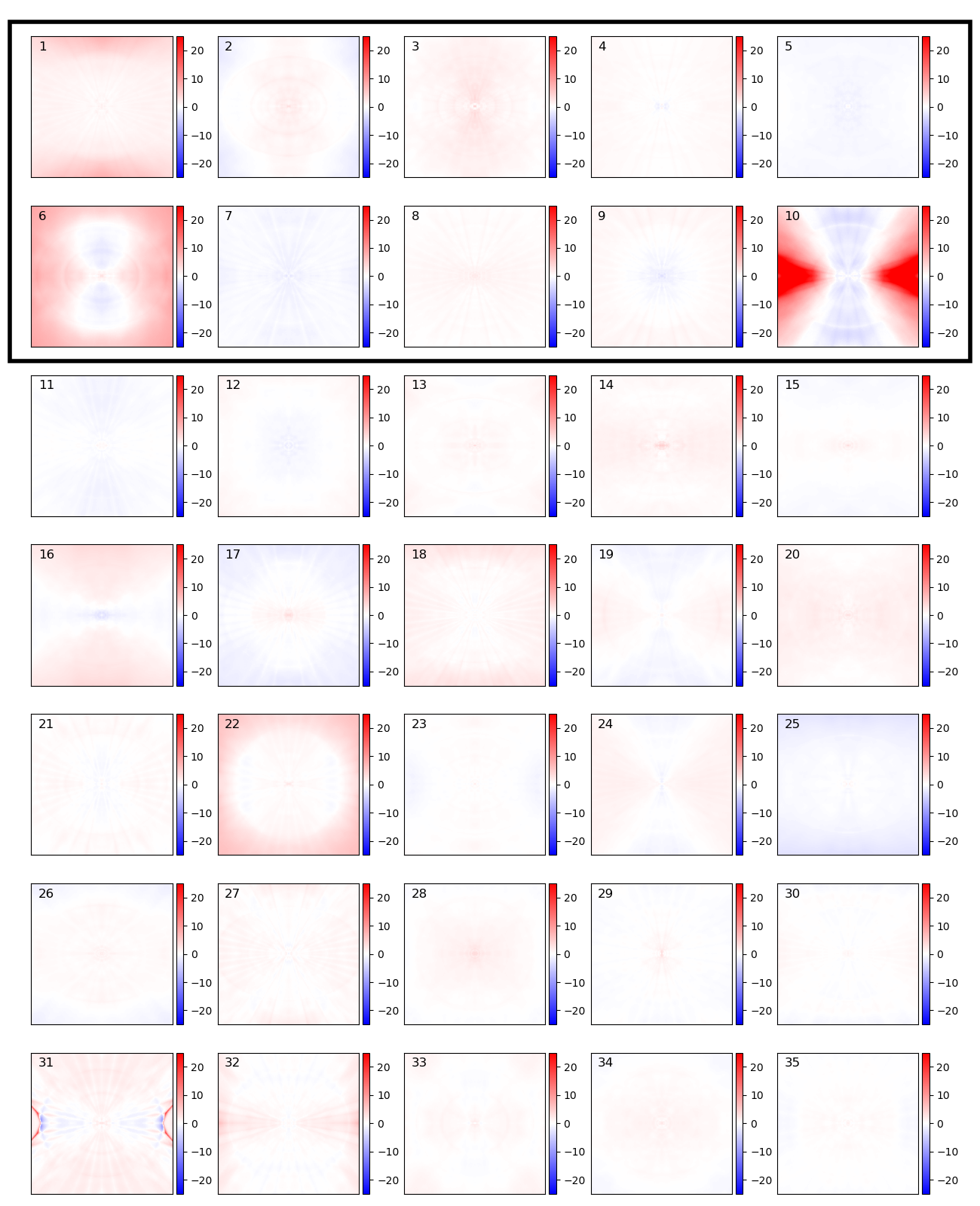}
    \caption{Example residuals (SKiNN prediction$-$JAM truth, km s$^{-1}$) are shown  for several randomly selected systems. The first 10 systems (boxed in black) are later used for joint inference in Section \ref{sec:joint_modeling}. The third and tenth systems are those from Fig. \ref{fig:goodbadresid}.}
    \label{fig:test_residuals}
\end{figure}

We gauge SKiNN's accuracy using the test set. Figure \ref{fig:goodbadresid} shows both a typical example emulation of a $v_{\rm rms}$ image (top row) and an example of a poor emulation (bottom row) which we consider to be a worst-case scenario. This scenario can arise in rare cases because for some parts of the parameter space, there can be segments of the map where the JAM truth has pixel values of 0 km s$^{-1}$. We discuss this case in more detail below. Because of this, relative residuals are often not the best metric to quantify error, and so instead we will use the non-relative error (third column in Fig. \ref{fig:goodbadresid}) to quantify our performance. Figure \ref{fig:test_residuals} shows the residuals for 35 systems. We have selected the first 10 systems, including our worst-case system (boxed in black in Fig. \ref{fig:test_residuals}) to explore in the context of joint lensing+kinematics inference in Section \ref{sec:joint_modeling}, while the remaining 25 examples are randomly selected from the test set. The 31st (bottom left) system in Fig. \ref{fig:test_residuals} is another example with zero-valued pixels, albeit a less egregious case than our worst-case example. 

\begin{figure}
    \centering 
    \includegraphics[width=0.95\linewidth]{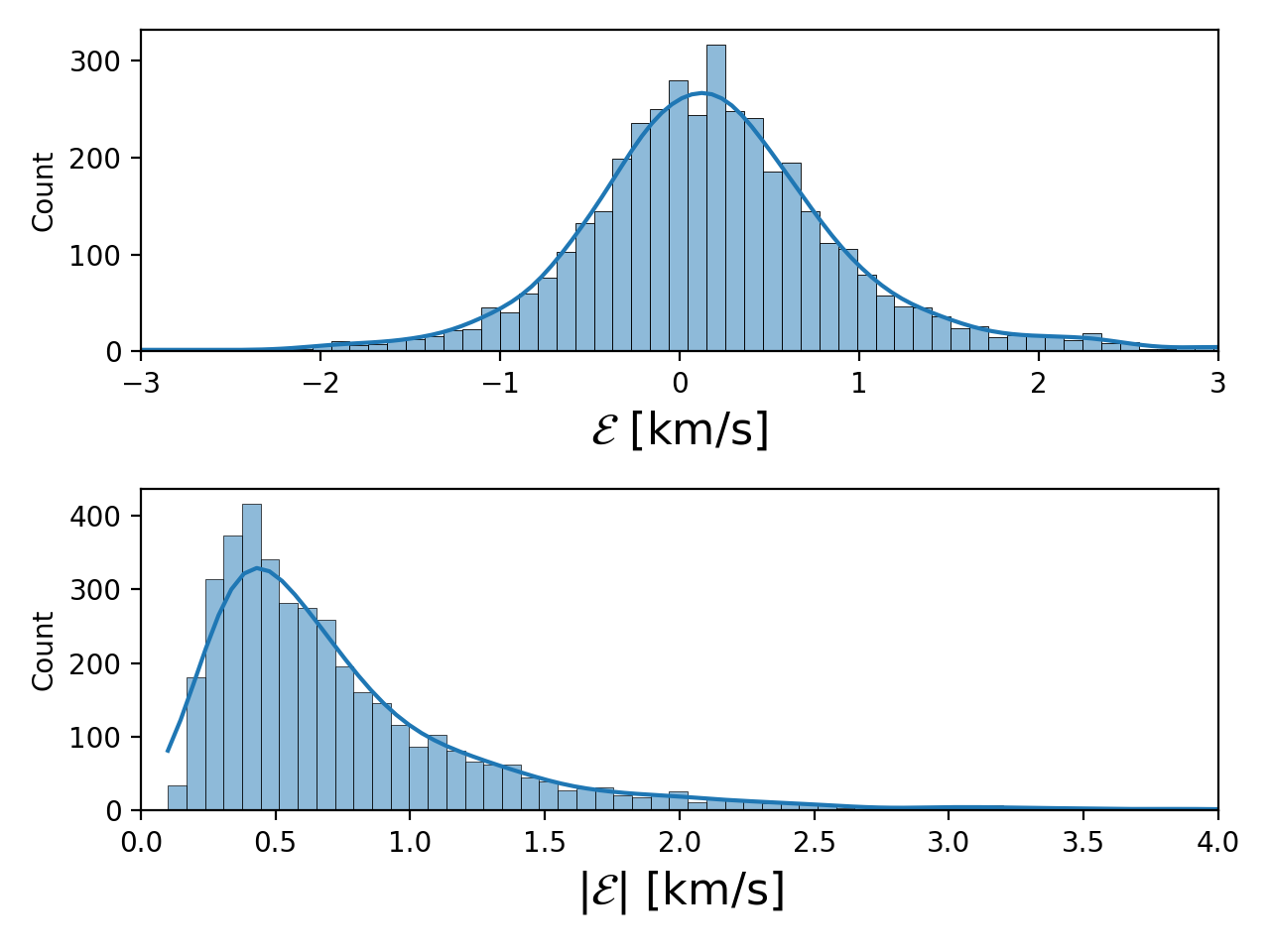}
    \caption{Error evaluated as (prediction$-$truth), averaged across each $v_{\rm rms}$ image in the test set. Top: mean error averaged over each image (Eq. \ref{eq:mean_error}), bottom: mean absolute error averaged over each image (Eq. \ref{eq:mean_abs_error}) }
    \label{fig:test_error}
\end{figure}

Aside from the two instances with zero-valued pixels, every single pixel across the remaining 33 images has an error of less than 10\,km s$^{-1}$, with typical error significantly less than that. To give an estimate of the accuracy of each image, we calculate the error averaged over each image,
\begin{equation}\label{eq:mean_error}
    \mathcal{E} = \frac{1}{d^2}\sum^{d^{2}} \Psi_{\theta}(\mathbf{x})-\mathbf{y},
\end{equation}
as well as the absolute error averaged over each image,
\begin{equation}\label{eq:mean_abs_error}
    |\mathcal{E}| = \frac{1}{d^2}\sum^{d^{2}} |\Psi_{\theta}(\mathbf{x})-\mathbf{y}|.
\end{equation}
These measures give an idea of the average error of a typical pixel across an image. We plot these measures for each image in our set of 4000 test images in Fig. \ref{fig:test_error}. Most systems are very well emulated: the mean error averaged over each image $\mathcal{E}$ is centered on 0.16\,km s$^{-1}$, with a spread of approximately 0.88\,km s$^{-1}$, indicating that SKiNN does not introduce a bias by overpredicting or underpredicting the image values. The image-averaged absolute error $|\mathcal{E}|$ has a median of $0.6$\,km s$^{-1}$, with the 95th percentile corresponding to an error of 2\,km s$^{-1}$. We find that the distributions of these measures of error are quite similar when considering only the pixels in the innermost 2\arcsec\ (black circles in Fig. \ref{fig:goodbadresid}), indicating that the central region, where real data is most sensitive, is equally well-emulated. With the knowledge that typical maps of lensing ETGs have $v_{\rm rms}\sim 200$\,km s$^{-1}$, this indicates the emulated images are accurate to better than $1\%$ in nearly all cases. Considering that real observations of $v_{\rm rms}$ have an approximate precision of $6-7$\,km s$^{-1}$ \citep{Cappallari11}, we consider this an excellent emulation of JAM.

\begin{figure}
    \centering 
    \includegraphics[width=0.95\linewidth]{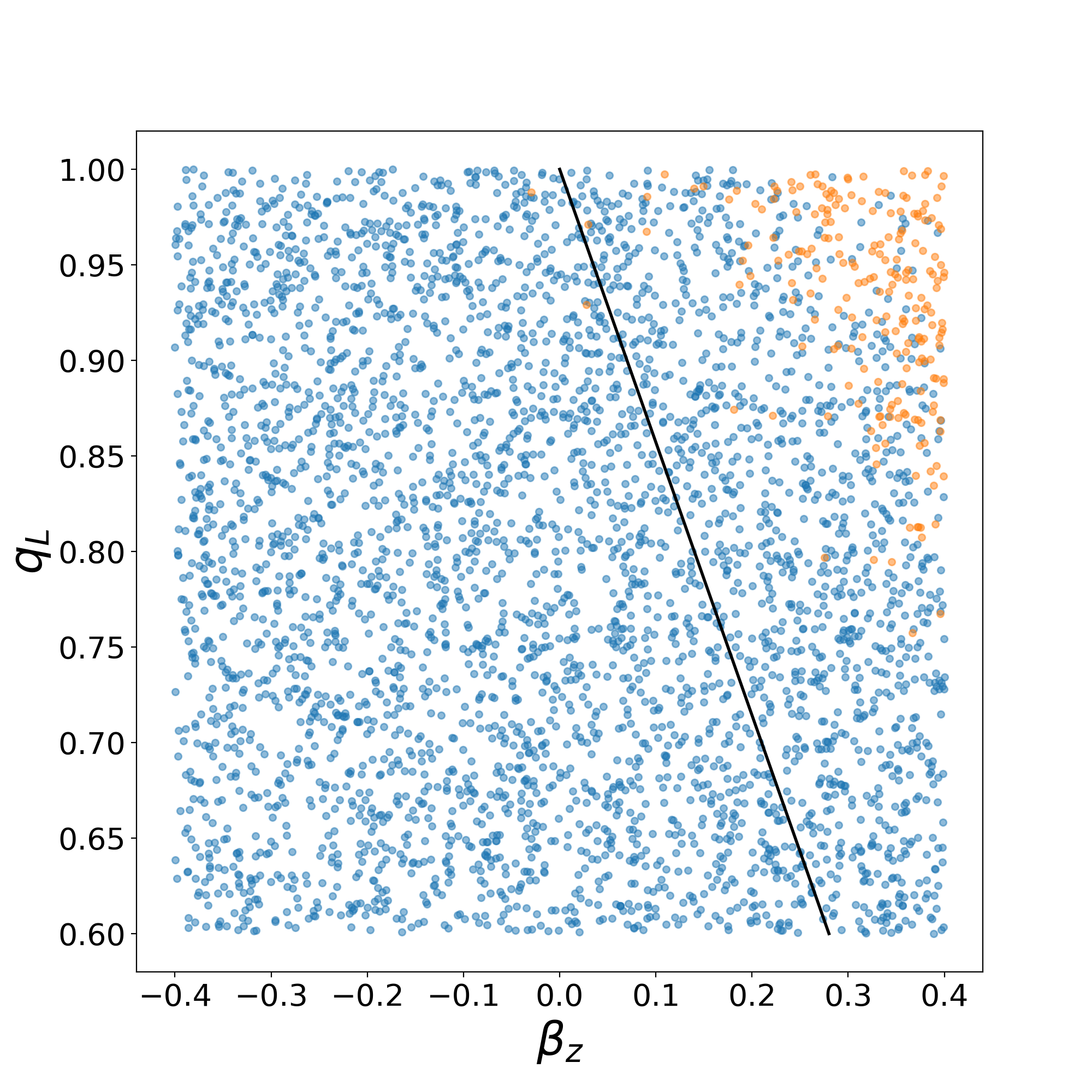}
    \caption{Scatter plot where each point represents the input parameters for each of the 4000 test-set images. Maps with zero-valued pixels (orange) are predominately confined to a small region of parameter space with high $\beta_z$ and $q_{\rm L}$, compared to those with nonzero pixel values (blue). The black line indicates the cutoff of \citet{Cappellari16}, above which ETGs have not been observed.}
    \label{fig:zero_pix}
\end{figure}

While most systems are very well emulated, there is a tail to the absolute error distribution, and so we looked for the systems with the highest error to see if we could diagnose any weaknesses in the emulation. This led us to discover systems like the worst-case system in the bottom row of Fig. \ref{fig:goodbadresid}, where the truth maps have pixels with a value of zero in the outer regions. These systems comprise approximately 5\% of the training/test sets. In the inner regions of the maps, these systems still provide quite good emulations, even in the worst case, indicating that the NN has learned well how to handle these inner features. Diagnosing the cause of the zero-valued pixels, we find that these cases are confined to a region of parameter space with high $\beta_z$ and high $q_{\rm L}$, as shown in Fig. \ref{fig:zero_pix}. This region of parameter space is unlikely to be relevant for real ETGs, which typically have $\beta_z<0.7\epsilon$ where $\epsilon=1-q$ \citep{Cappellari16}. SKiNN is less successful with fitting these systems than with the full set, likely because the sharp dropoff to zero must be captured precisely to avoid substantial residuals. The absolute error for these systems is about twice that of the full set, with median 1.4\,km s$^{-1}$ and 95th percentile of 3.9\,km s$^{-1}$.

We visually inspect the mean absolute error over the whole parameter space to check to see if there are any other regions in which SKiNN performs better or worse. We find that SKiNN performs slightly worse than average for systems with high $\beta_z$ or low $\gamma$. More quantitatively, in the regions where $\beta_z>0.3$ or $\gamma<1.6$, the median absolute error is approximately 0.9\,km s$^{-1}$ with 95th percentile of 3.1\,km s$^{-1}$. These regions are also unlikely to be sampled by realistic galaxies based on the $\beta_z$ constraints of \citet{Cappellari16} and the fact that ETGs have nearly isothermal slopes \citep{Shajib19a,Shajib21}.

Altogether, SKiNN emulates JAM to within $\sim1$\,km s$^{-1}$ in most cases, with some parts of the parameter space having errors in the recreation $\sim5$\,km s$^{-1}$. While this is still sufficient for most applications, we note that a user can always use SKiNN to find an approximate solution and afterwards use JAM to confirm the accuracy of converged minimum, which still saves considerable computational time. In Section \ref{sec:joint_modeling}, we will show that using SKiNN to recover the input parameters results in an accurate constraint for the 10 systems in the black box in Fig. \ref{fig:test_residuals} shown above, including the worst-case scenario with zero-valued pixels.

\subsection{Speed}

Using the same machine as before, emulation of a $v_{\rm rms}$ image with SKiNN takes approximately 100\,ms using an NVIDIA GeForce GTX 1660 Super GPU, which is about 200 times faster than generating the same image using JAM. This does require the upfront cost to build the training and test sets, which took approximately 2.5 days, as well as the training cost, which took approximately 3 days. However, these costs only need to be performed once, and have already been done for the current setup, for which we make the trained weights publicly available as part of the SKiNN python package. This speed increase makes it possible to sample $v_{\rm rms}$ kinematics within an MCMC, which we implement and demonstrate in the following section.

\section{Joint inference} \label{sec:joint_modeling}
In a joint implementation, the lens light and mass models are used to generate models of both the imaging data and binned kinematic data which are respectively compared to the observed image of the lensing system and the kinematics of the lensing galaxy (Fig. \ref{fig:SKiNN_scheme}). Each comparison leads to a likelihood estimation, and the summed log likelihood is optimized, hence constraining the lens mass and light parameters by using both imaging and kinematic data. Some parameters, such as the source parameters (not shown in this work), are constrained only from lensing. On the other hand, some parameters, such as the inclination and anisotropy, are specific to the kinematic data and cannot be constrained by the lensing imaging information. Meanwhile, the mass profile and lens light profile are constrained by both types of data, reducing modeling degeneracies.

\subsection{Kinematic likelihood}
In this section, we describe the steps needed to transform the output from SKiNN into a binned kinematic data and evaluate a likelihood. SKiNN outputs a high-resolution $v_{\rm rms}$ image at a fiducial cosmological distance, which must be rescaled, rotated, sampled, and binned to compare with the observed kinematic data.

The first step necessary is to rescale the map from the fiducial cosmological distances to those of the system. Rescaling with cosmological distance is a straightforward multiplicative factor on the values of each $v_{\rm rms}$ pixel \citep{Birrer20, Yildirim23}, scaling as 
\begin{equation}\label{eq:distance_scaling}
    v_{\rm rms} \propto \sqrt{\frac{D_{\Delta t}}{D_{\rm d}(1+z_{\rm d})}},
\end{equation}
where $D_{\rm d}$ is the distance to the deflector, $z_{\rm d}$ is the deflector redshift, and the time-delay distance $D_{\Delta t}=(1+z_{\rm d}) \frac{D_{\rm d} D_{\rm s}}{D_{\rm ds}}$ can be expressed in terms of the distance to the source $D_{\rm s}$ and the distance between the deflector and source $D_{\rm ds}$. In a joint framework, $D_{\Delta t}$ is constrained by the lensing model of imaging and time delays, allowing the kinematics to constrain $D_{\rm d}$ through Eq. \ref{eq:distance_scaling}. Our training set uses fiducial distances corresponding to a lens redshift of $z_{\rm d}=0.5$ and source redshift of $z_{\rm s}=2$ for a flat universe with $H_0=72 \text{ km s}^{-1} \text{Mpc}^{-1}$ and $\Omega_{\rm m}=0.32$, resulting in $D_{\rm d}=1216$ Mpc and $D_{\Delta t}=2887$ Mpc.

Once rescaled according to the lens distances, the SKiNN image must be resampled to match the observed data. SKiNN-generated images are originally aligned with the x-axis centered on the origin, and as such must be first rotated and translated to align with the light distribution before binning down to the lower resolution of the observed data. Finally, the generated image is binned according to the same bins as the observed data, which uses Voronoi binning \citep{Cappellari03} to construct bins of approximately the same signal to noise ratio (S/N). In each bin, the luminosity-weighted $v_{\rm rms}$ is calculated using the modeled light distribution. We are left with a list of predicted $v_{\rm pred}$ values in each of the data bins, which can be compared with observed $v_{\rm data}$  to evaluate a likelihood, which to within a normalization constant is expressed as:
\begin{equation}\label{eq:chi2_kin}
    \log \mathcal{L_{\rm kin}} = -\frac12 \chi^2_{\rm kin}= -\frac12 (v_{\rm pred}-v_{\rm data})^\top \mathcal{C}^{-1} (v_{\rm pred}-v_{\rm data}),
\end{equation}
where $\mathcal{C}$ is the covariance matrix giving the precision to which $v_{\rm rms}$ can be measured in a each bin. This likelihood term is added to the lensing likelihood to construct a joint likelihood which can be maximized to recover a best-fit model of the lensing+kinematic data.

We have implemented this joint likelihood in \texttt{lenstronomy}, which is already capable of recovering a likelihood of a lens model given imaging data. The addition of SKiNN allows \texttt{lenstronomy} to evaluate a kinematic likelihood by using SKiNN to construct a $v_{\rm rms}$ map from the 8 parameters in $\mathbf{x}$ and then applying the rescaling, translation, and rotation to allow for sampling over the center position, position angle, and cosmological distances. The 5 parameters which dictate this transformation are listed in Table \ref{table:postNNparams}. All in all, a given map from which a kinematic likelihood can be evaluated is produced using the 13 parameters in Tables \ref{table:parambounds} and \ref{table:postNNparams}.

\subsection{Testing the joint inference framework}
To test the joint implementation of lensing+SKiNN, we create mock lensing and kinematic data which we fit both separately and jointly to demonstrate the utility of SKiNN in a realistic test case. 

\begin{figure} 
\includegraphics[width=0.9\linewidth]{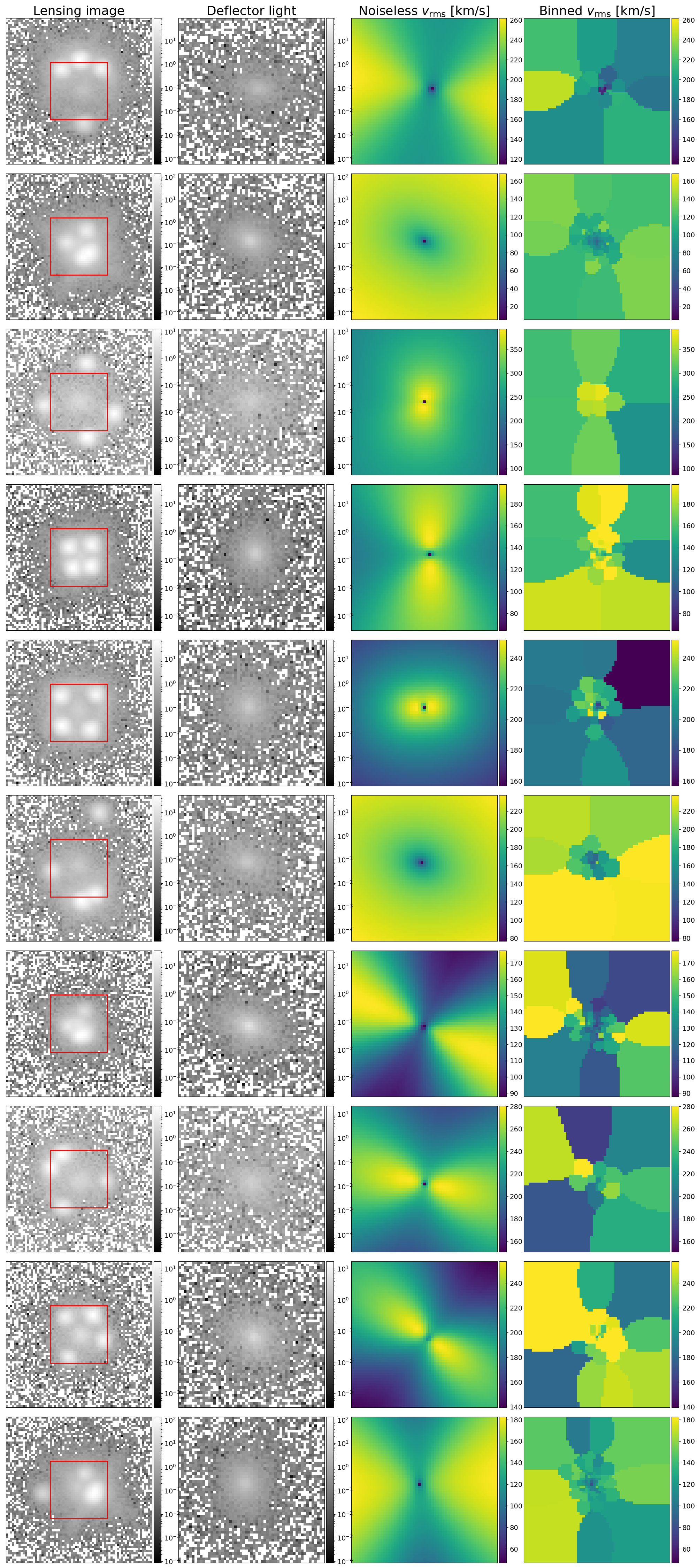} 
\caption{Mock observations for the 10 systems used for joint inference. From left to right: the lensing image used for lens modeling (in combination with time delays); the deflector light only, with the lensing arcs removed, used to define the kinematics binning and weighting; the noiseless $v_{\rm rms}$ map created by JAM, translated, rotated, and sampled at the data resolution; and the binned $v_{\rm rms}$ with noise added, i.e. the input for the kinematics modeling. The red square in the lensing imaging indicates the tighter field of view of the remaining three panels.}
\label{fig:mock_obs}
\end{figure}

We create 10 sets of mock imaging and kinematic data from the boxed systems in Fig \ref{fig:test_residuals}. These systems were randomly selected, except for the 10th system, which as discussed in Section \ref{sec:performance} was selected specifically to gauge our accuracy in the worst-case scenario. For a given system, a member of the test set provides the truth velocity map and the 8 parameters in Table \ref{table:parambounds}. We then randomly draw truth values for $\phi$, $x_{\rm center}$, $y_{\rm center}$, $D_{\Delta t}$, and $D_{\rm d}$ according to the ranges in Table \ref{table:postNNparams}.  We show these mock observations in Fig. \ref{fig:mock_obs}. Mock lensing imaging information is shown on the left. The binned $v_{\rm rms}$ image (far right) is constructed using the deflector light as a weight map for the unbinned $v_{\rm rms}$, which itself comes from JAM but has been rotated and rescaled according to the truth distance values for a given system. The remainder of this section discusses the creation of these mock data in more detail.

\begin{table}
\centering
    \renewcommand{\arraystretch}{1.3} 

    \begin{tabular}{c | c | c }
        
         Parameter & Description & Sample range\\
         \hline
         $\phi$ & Position angle & $[0,90\degree]$\\
         $x_{\rm center}$ & $x$-coordinate of center & $[-0.15\arcsec, -0.15\arcsec]$\\
         $y_{\rm center}$ & $y$-coordinate of center & $[-0.15\arcsec, -0.15\arcsec]$\\
         $D_{\Delta t}$ & Time delay distance & 2887 Mpc $\pm20\%$\\
         $D_{\rm d}$ & \multirow{2}{*}{\shortstack[c]{Angular distance\\to the deflector}}  & 1216 Mpc $\pm20\%$\\
         \cr

    \end{tabular}
     \caption{Parameters describing the translation, rotation, and rescaling from a SKiNN output map to data resolution. Parameters are sampled uniformly within the range indicated when constructing mock data. }
     \label{table:postNNparams}
\end{table}

For the mock kinematic data, a light map is needed to define the binning scheme. We construct a  mock image of the S\'ersic light distribution, where the goal of this mock is to have a realistic binning scheme with $\sim 20-50$ bins to test the SKiNN implementation rather than to perfectly calibrate to the real noise levels of any particular telescope. That said, our brightness and noise settings correspond to setting the integrated brightness to a magnitude 19 galaxy with a $200$s exposure time using a zero point of 26 magnitude and read out noise of 21\,${\rm e}^{-1}/s$ with the magnitude of the sky background set to 20 mag. The $v_{\rm rms}$ mock image is created using a $55\times55$ pixel grid with a resolution of 0.05\arcsec. We evaluate the luminosity-weighted $v_{\rm rms}$ using the test image, which was created using JAM. To mimic the effects of atmospheric seeing, we convolve the high-resolution JAM $v_{\rm rms}$ as well as the light during the weighting step using a Gaussian PSF with a FWHM of 0.1\arcsec~ before binning. We apply Voronoi binning to the data with a target S/N $\sim15$ for each bin. Finally, for each $v_{\rm rms}$ bin, we independently add noise drawn from a Gaussian distribution with a width depending on the bin S/N: the width is set to $10\%$ at S/N $=10$ and narrows with increasing S/N to a minimum width of $5\%$ when the bin S/N $\ge40$. This scatter added to the $v_{\rm rms}$ values is intended to represent the imperfect accuracy to which the velocity can be measured from the spectrum of a given bin. The covariance matrix in Eq. \ref{eq:chi2_kin} is therefore a diagonal matrix where the entry for each bin is the variance of each Gaussian. This binned data is taken to be our observed $v_{\rm rms}$ data, shown in the far right column of Fig. \ref{fig:mock_obs}.

The lensing imaging data is constructed in a similar fashion to the previously mentioned mock light map used for the kinematics weighting. We use the same exposure times and noise settings, but add in arcs from a lensed source. To create these arcs, we randomly draw a source position from within the region of the source plane capable of producing four lensed images. In this position, we place a point source with an intrinsic magnitude of 21.75 in the center of an extended circular S\'ersic source with $R_{\rm Sersic}=0.1$ and intrinsic magnitude of 22.5.  Unlike the kinematic light map, we change the image cutout to a $70\times70$ pixel cutout with a resolution of 0.1\arcsec. Since the lensing imaging includes point sources, we use the PSF from the TDLMC \citep{Ding21}, which is more realistic than a Gaussian PSF for imaging data. These changes  allow for a larger cutout than the velocity map to ensure that all the lensed images are observed. 

In addition to the imaging information, the lens data include mock relative time-delay measurements of the multiple images. These measurements are given a relative uncertainty of $2\%$ or 1 day, whichever is higher, and are used to evaluate a time delay likelihood which is added into the total likelihood to sample $D_{\Delta t}$.

In the joint framework, the kinematic data and the lens imaging data are combined together along with the time delay data to optimize a total log likelihood by adding their respective log likelihoods (i.e., treating observations as independent and multiplying their likelihoods), which to within the normalization of $L$ is equivalent to a summed $\chi^2$: 
\begin{equation}\label{eq:logL}
\log L_{\rm tot}= -\frac12 \chi^2_{\rm tot}= -\frac12 \left(\chi^2_{\rm img}+\chi^2_{\rm kin}+\chi^2_{\rm TD}\right).
\end{equation}
However, we note that the log likelihood of lens imaging data is preponderant over the kinematic and time-delay likelihoods. Indeed, images encompass a lot of pixels ( $\sim5000$) while the number of kinematic bins is restricted to the order of $\sim30$ and the time delay likelihood is restricted to $3$ observations at most. When images are taken with the best telescopes, that is with higher resolution, the relative contribution of the imaging data gets even higher \citep[e.g., TDCOSMO systems;][]{Millon20}. The optimal way to combine likelihoods of different magnitudes may require rescaling individual component likelihoods and is a general problem which is beyond the scope of this work. In our case, we simply combine likelihoods with no rescaling as in Eq. \ref{eq:logL} and show that the inclusion of kinematics still improves the overall constraint.

\subsection{Results of the joint framework test}

For each of the systems in Fig. \ref{fig:mock_obs}, we model the lensing image and the kinematic image individually as well as jointly. 

The lens modeling is performed on the mock images and time delays using a PEMD+external shear model. Strictly speaking this is more optimistic for a lens model than we should expect from real systems because we know in this case that the true mass distribution is the same as our model, meaning we will artificially break the MSD by giving the lens model the correct profile. As such, our lensing inference will reflect a precision for the slope $\gamma$ which is overestimated relative to a more realistic case. Nonetheless we will show that the kinematic information helps constrain $\gamma$ to break the MSD in the more natural way by measuring the mass distribution directly.

In all cases, we use an MCMC to sample the posterior parameter space near the truth label. This assumes that a blind inference would find a maximum likelihood for parameters near the truth value from which to start an MCMC, which may not always be true. However, seeing as we ultimately find that the MCMC converges to a region surrounding the truth value, we are satisfied that this maximum would be recovered from a blind starting point, and felt our computational resources were better spent on exploring more systems than on converging from an arbitrary starting point.

\begin{figure*} 
\begin{tabular}{cc}
\includegraphics[width=\textwidth]{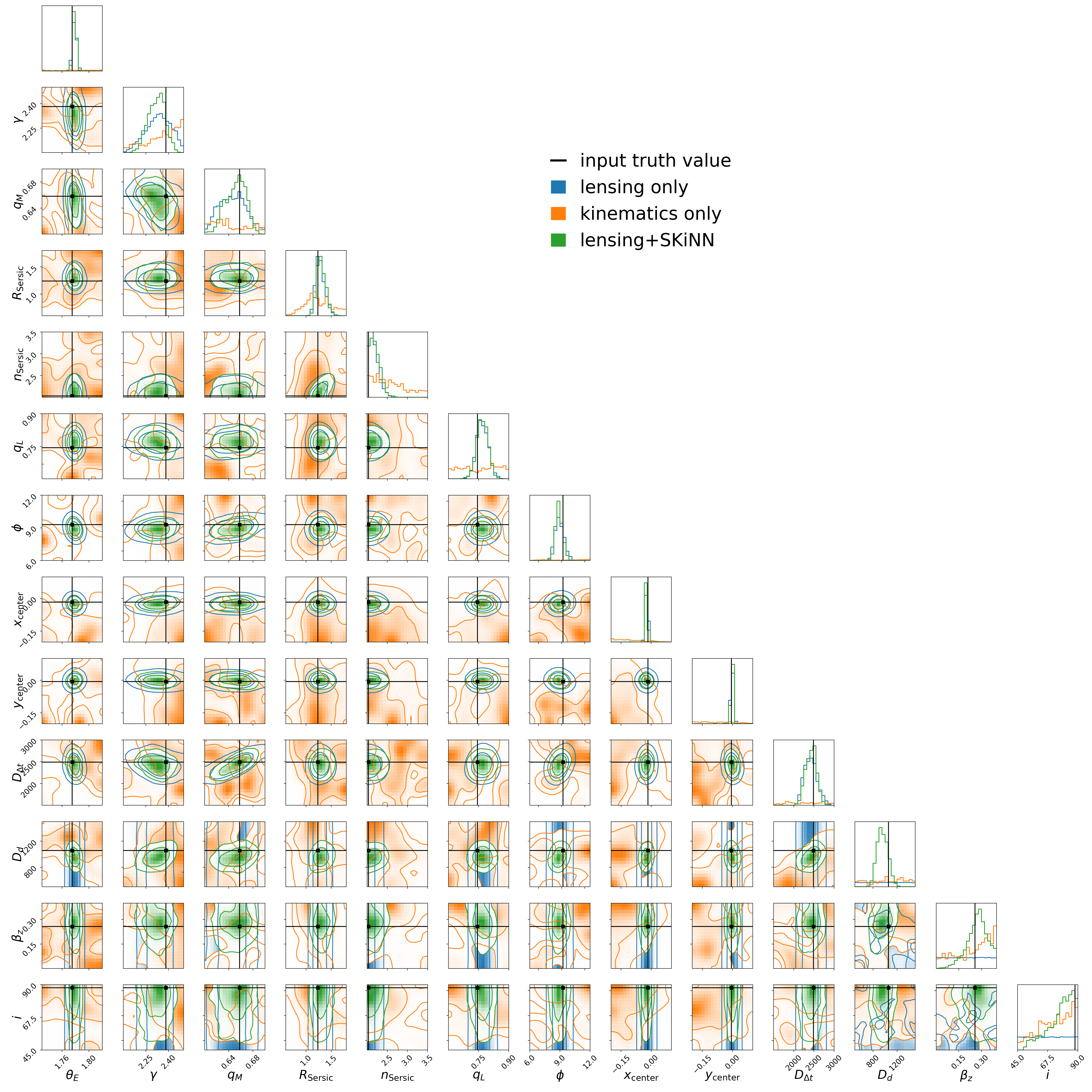} \\
\end{tabular}
\caption{Corner plot for the MCMC sampling of input parameters using SKiNN for the first example in Fig. \ref{fig:goodbadresid}.}
\label{fig:joint_cornerplot}
\end{figure*}

We plot the MCMC results for one system in Fig. \ref{fig:joint_cornerplot}. Sampling using the lensing data alone (blue) provides constraints on most parameters, with no ability to constrain $D_{\rm d}$, $\beta_z$, or $i$, since these three parameters are not sampled in a lensing-only model. As such, we have replaced them with uniform distributions for visualization in Fig. \ref{fig:joint_cornerplot}. The kinematics-only result (orange) typically cannot constrain individual parameters such as $\theta_{\rm E}$ as tightly as the lensing information can, but it is able to probe all of the parameters of interest. When sampled jointly (green), the resulting posteriors narrow around the truth values, indicating the kinematics constraints have helped the lensing constraints inform the mass distribution.

\begin{figure*} 
\begin{tabular}{cc}
    \includegraphics[width=\textwidth]{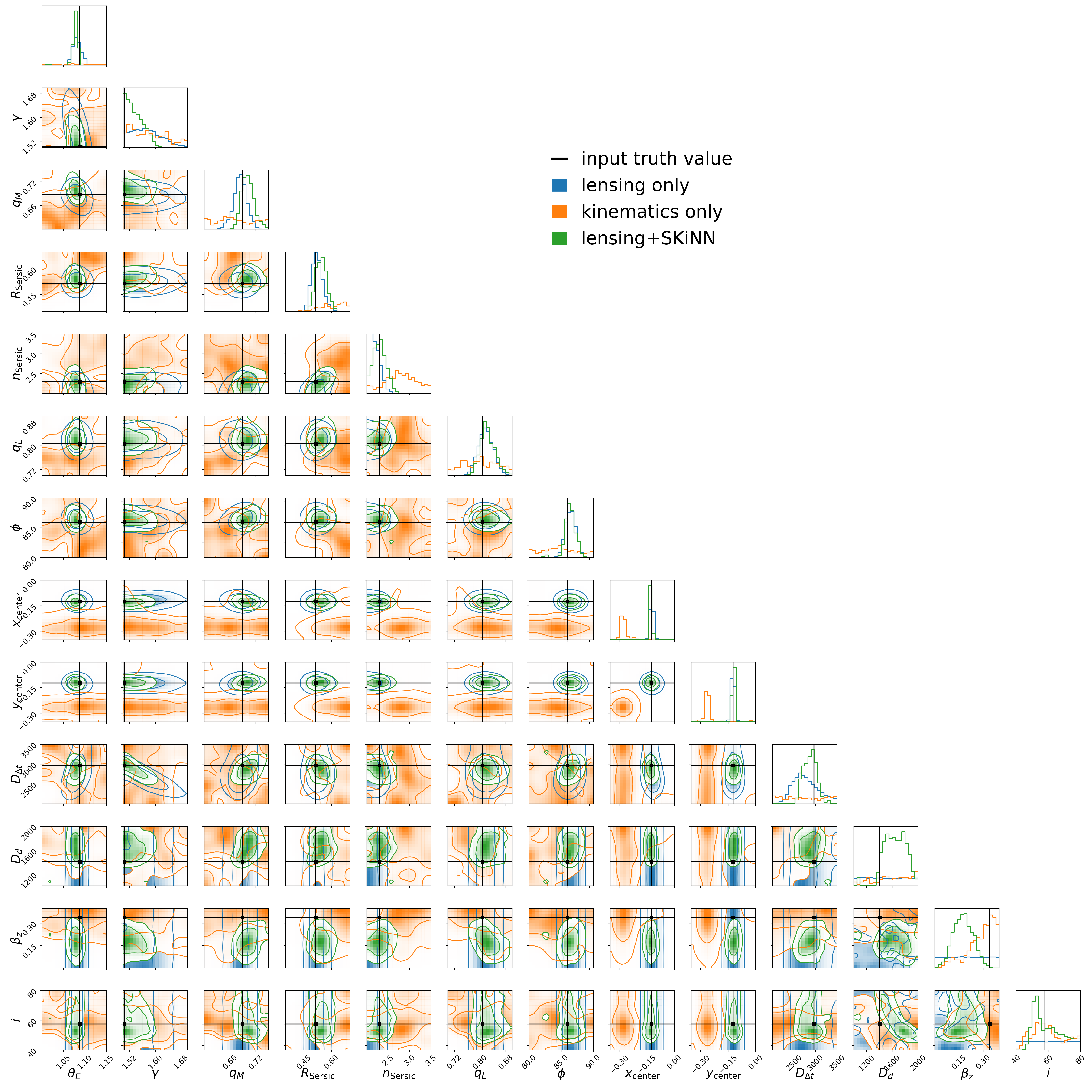} \\
\end{tabular}
\caption{Corner plot for the MCMC sampling of input parameters using SKiNN for the worst-case system with zero-valued pixels (second example in Fig. \ref{fig:goodbadresid}).}
\label{fig:cornerplot_zeropix}
\end{figure*}

In Fig. \ref{fig:cornerplot_zeropix} we plot the results of the MCMC for the worst-case scenario system with zero-valued pixels. The results are quite similar, possibly because the weighting of the light distribution favors the central regions, as represented by the black circle in Fig. \ref{fig:goodbadresid}. This makes the outer regions where numerical effects are present less relevant. One possible bias in the kinematics-only result is that the truth centroid position ($x_{\rm center}$ and $y_{\rm center}$) lies just outside the $\sim1\sigma$ level of the posterior. Fortunately the centroid positions are well-constrained by imaging information, and the joint inference strongly favors the correct centroid positions. Noting this, one could also consider for a specific modeling context to fix the centroid positions to the values determined by the lens model. For our tests, we find that fixing these centroid positions does not significantly change the other parameter posterior distributions. 

\section{Discussion} \label{sec:discussion}
We have shown that SKiNN offers a fast way to emulate JAM to high accuracy. The speed increase has made it possible to jointly model lensing and kinematics for several system on a single GPU. Here we summarize the joint inference results of the 10 systems and discuss our outlook for SKiNN in the future. 

\begin{figure} 
\includegraphics[width=\linewidth]{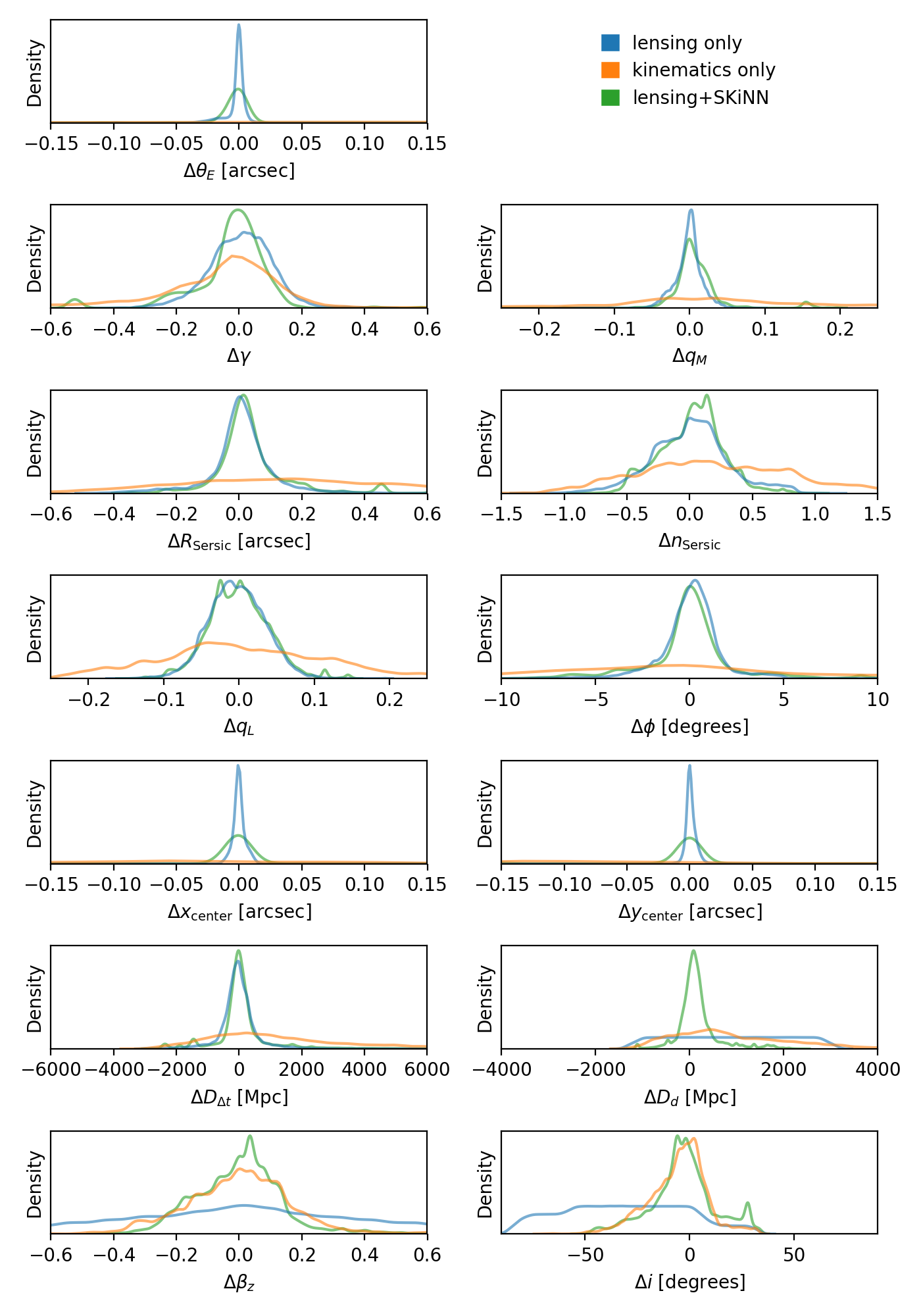}
\caption{The 1D distributions showing the recovery of the 13 parameters from Tables \ref{table:parambounds} and \ref{table:postNNparams}. Plotted results are stacked from those of the 10 systems in Sec. \ref{sec:joint_modeling}. }
\label{fig:1Dhistograms}
\end{figure}

Seeing as the full cornerplots are somewhat cumbersome to show for all 10 systems, we show the 1D histograms for the 13 parameters, normalized to the truth values in Fig. \ref{fig:1Dhistograms}. These plots are constructed by stacking together all the chains of $\Delta x=(x-{\rm truth})$ for each parameter $x$, and plotting all 10 systems as one distribution. While this visualization can be lacking for parameters with truth values near the edge of the prior,
    \footnote{This can be seen most easily from the lensing-only distributions of $D_{\Delta t}, \beta_z$, and $i$, which were uniform distributions in the corner plots, but when stacked together can result in non-centered distributions, since for example $D_{\Delta t}$ cannot go below zero so it is possible to be considerably off in the positive direction but impossible to as offset in the negative direction.}
it serves to visualize the precision with which data of this quality is able to constrain the truth.

A caveat that bears repeating is that since this test used a PEMD lens model to fit a PEMD mass distribution, the MSD is artifically broken due to having the external knowledge of the functional form of the mass model. As such, the uncertainty of the lensing-only $\gamma$ result is underestimated. This makes it all the more important that the kinematics-only $\gamma$ result be centered on the truth to show that the kinematics helps to break this degeneracy, which is the result we find in Fig. \ref{fig:1Dhistograms}.

From the plots, it is clear that most parameters are constrained by lensing alone, with the inclusion of kinematics helping only slightly. Some parameters are constrained even better by lensing alone than by a joint inference such that it may be preferable to simply fix them to the lensing result: namely $\theta_E$, $x_{\rm center}$, and $y_{\rm center}$. Meanwhile the anisotropy and inclination are mostly informed by the kinematics, with the lensing helping only slightly. The notable exception to systems being constrained predominately by one form of information or the other is $D_{\rm d}$, which clearly requires joint information to constrain. 
The kinematic constraint can only recover a combination of $D_{\rm d}$ and $D_{\Delta t}$, a degeneracy which is broken by the lensing measure of $D_{\Delta t}$. The precise constraint of these cosmological distances is critical for a reliable recovery of $H_0$. We show that for all systems the combination of lensing and kinematic data is consistently able to break the degeneracy and recover accurate cosmological distances. 

Despite the significant speed increase SKiNN has over JAM, it is still the bottleneck, taking approximately 90\% of the time for a each likelihood evaluation. A typical joint fit in this test undergoes nearly $10^6$ likelihood evaluations, each of which takes approximately 100\,ms using a single GPU ($\sim28$ GPU-hours per system). Note, however that if the same test were done using JAM, we estimate it would recover the same results but take $\sim 5800$ CPU-hours per system, a prohibitive cost for modelers without the use of cluster computing.

\subsection{Limitations and Outlook}\label{sec:limitations}

The application of SKiNN in its current form is limited to the assumptions used to create the training set. Constructed with lensed quasar systems in mind, the training set used in this work may not be suitable for all applications, depending on the expected range of parameter values (for example demanding that $R_{\rm Sersic}<\theta_E$, which is not always true for lower redshift galaxy-galaxy lenses). Being an emulation of JAM, SKiNN inherits JAM's limitations: the mass is assumed to be oblate, axisymmetric, and with a deprojectable axis ratio. SKiNN is limited at present to a constant-anisotropy power-law mass model and a single Sersic light profile, with the position angles and centroids aligned. 

We anticipate that generalization of the SKiNN method is possible, as we have previously generalized from an isothermal prototype to the power-law mass profile discussed in this work. Such generalization requires recreating (or supplementing) the existing training set and retraining the neural network. For example, one could expand the method beyond the limitations of the JAM model if one had access to a large training set created from N-body simulations or some other method of emulating galaxy kinematics. In such a case, one could speed up training time by using transfer learning starting from the weights learned in this work. 

Thinking more broadly, it may even be possible to modify the method of SKiNN to input a mass and light map instead of a parametric description, providing great flexibility, but this would require an update to the architecture as well as retraining, and is beyond the scope of this current work. However, there is reason to be optimistic about such modifications: we have already iterated on the design of SKiNN from the NeurIPS version \citep{NeurIPS_version} to a new architecture and a larger training set, indicating that similar iterations are possible in the future.

\section{Conclusion}\label{sec:conclusion}

We present the Stellar Kinematics Neural Network (SKiNN), which emulates $v_{\rm rms}$ maps for dynamical modeling in the context of joint modeling with strong gravitational lensing. SKiNN is trained to emulate Jeans Anisotropic MGE (JAM) dynamical
modeling, which it does to high precision (better than $1\%$ in nearly all cases), with no indications of a bias. SKiNN makes it possible to compute kinematic likelihoods approximately 200 times faster than with JAM. This speedup reduces the severity of the computational bottleneck that makes joint kinematic+lensing inference expensive, allowing us to sample MCMC chains for a kinematic likelihood in a timely manner on a single GPU. We show that these sampling methods are able to recover the input parameters associated with the truth when provided mock data using JAM. 

SKiNN is currently available for use as a python package and ready to use for time-delay cosmography applications. While SKiNN is currently implemented in \texttt{lenstronomy}, its modular nature makes it suitable to implement in other lens modeling codes. SKiNN is fully differentiable, and so its value can be further optimized if used in conjunction with differentiable modeling software. With updates to the training set and/or architecture, the method of SKiNN can likely be further generalized to a wider range of mass and light profiles.

This work represents a step forward in making modern dynamical methodology tractable for strong lens modeling, increasing the complexity from spherical Jeans to axisymmetric Jeans. This increased model complexity allows lens modelers to make proper use of upcoming spatially resolved kinematics from modern JWST-era telescopes. 

\begin{acknowledgements}
This work originated in the Lensing Odyssey 2021 workshop, and so we would like to acknowledge the organizers and attendees for the fruitful discussions. 
This project has received funding from the European Research Council (ERC) under the European Union’s Horizon 2020 research and innovation programme (COSMICLENS: grant agreement No 787886; LENSNOVA grant agreement No 771776). SE and SHS thank the Max Planck Society for support through the Max Planck Research Group and Max Planck Fellowship for SHS. This research is supported in part by the Excellence Cluster ORIGINS which is funded by the Deutsche Forschungsgemeinschaft (DFG, German Research Foundation) under Germany’s Excellence Strategy -- EXC-2094 -- 390783311. AG thanks the Swiss National Science Foundation for support through the SNSF \textit{Postdoc.Mobility} programme. GV's research was made possible by the generosity of Eric and Wendy Schmidt by recommendation of the Schmidt Futures program.

\end{acknowledgements}


\bibliographystyle{aa}
\bibliography{biblio}






\end{document}